\documentstyle[11pt,moriond,epsfig]{article}

\def\be{\begin{equation}}
\def\ee{\end{equation}}
\def\bea{\begin{eqnarray}}
\def\eea{\end{eqnarray}}

\begin{document}
\vspace*{4cm}
\title{ANTIMATTER FROM PRIMORDIAL BLACK HOLES}

\author{Aur\'elien Barrau}

\address{Institut des Sciences Nucl\'eaires \& UJF\\
53, av des Martyrs, 38026 Grenoble cedex, France\\
{\rm XIVth RENCONTRES DE BLOIS, MATTER-ANTIMATTER ASYMMETRY}}

\maketitle

\abstracts{Antiprotons and antideuterons are considered as probes to look for primordial
black holes in our Galaxy. I give a brief overview of the latest developments on the 
subject.}

\section{Introduction}

Primordial black holes (PBHs) could have formed in the early universe from
the collapse of overdense regions through significant density fluctuations. 
Their detection nowadays is a great challenge as it could
allow both to check the Hawking evaporation mechanism and to probe the early universe 
on very small scales that remain totally out of the range investigated by CMB or LSS 
measurements. They have recently been searched by their gamma-ray radiation \cite{MacGibbon2}
\cite{Carr3}, extremely high-energy cosmic-ray emission \cite{Barrau3},
and antiproton emission \cite{Orito}. This brief paper gives the latest
improvements obtained with antiprotons and antideuterons. Such antinuclei are very
interesting as the background due to spallation of cosmic protons and helium nuclei on the
interstellar medium is expected to be very small.

\section{Source term}

The Hawking spectrum \cite{Hawk1} for
particles of energy $Q$ per unit of time $t$ is, for each degree of
freedom:
\begin{equation}
\frac{{\rm d}^2N}{{\rm d}Q{\rm
d}t}=\frac{\Gamma_s}{h\left(\exp\left(\frac{Q}{h\kappa/4\pi^2c}\right)-(-1)^{2s}\right)}
\end{equation}
where $\kappa$ is the surface 
gravity, $s$ is the
spin of the emitted species and
$\Gamma_s$ is the absorption probability.
As it was shown by MacGibbon and Webber \cite{MacGibbon1}, when the
black hole temperature is
greater than the quantum chromodynamics confinement scale
$\Lambda_{QCD}$, quarks and gluons jets are
emitted instead of composite hadrons. To evaluate the number of
emitted antiprotons $\bar{p}$ , one therefore
needs to perform the following convolution:
\begin{displaymath}
      \frac{{\rm d}^2N_{\bar{p}}}{{\rm d}E{\rm d}t}=
      \sum_j\int_{Q=E}^{\infty}\alpha_j\frac{\Gamma_j(Q,T)}{h}
      \left(e^{\frac{Q}{kT}}-(-1)^{2s_j}\right)^{-1}
      \times\frac{{\rm d}g_{j\bar{p}}(Q,E)}{{\rm d}E}{\rm d}Q
\end{displaymath}
where $\alpha_j$ is the number of degrees of freedom, $E$ is the
antiproton energy and
${\rm d}g_{j\bar{p}}(Q,E)/{\rm d}E$ is the normalized differential
fragmentation function, {\it i.e.}
the number of antiprotons between $E$ and $E+{\rm d}E$ created by a
parton jet of type $j$ and energy
$Q$ (including decay products). The fragmentation functions have been 
evaluated with the
high-energy physics event generator
{\sc pythia}/{\sc jetset} \cite{Tj},
based on the string fragmentation
model.\\

To evaluate the antideuterons production, we used a simple coalescence scheme implemented 
directly within the PBH jets. This approach is similar to
the one used in Chardonnet {\it et al.} \cite{chardonnet} and Donato {\it et al.}
\cite{fiorenza}. The hadron momenta
given by PYTHIA can be compared together and each time an antiproton and an
antineutron are found to lie within the same coalescence sphere, an
antideuteron is created. As the coalescence momentum $p_0$ is not Lorentz
invariant, the condition must be implemented in the correct frame, 
namely in the
antiproton-antineutron center of mass frame instead of the laboratory one. 
Depending on the models and experiments the value was found to vary 
between 60
MeV/c and 285 MeV/c.
The number of antideuterons therefore reads as

\begin{equation}
\frac{{\rm d}^2N_{\bar{D}}}{{\rm d}E{\rm d}t}=
\sum_j\int_{Q=E}^{\infty}\alpha_j\frac{\Gamma_{s_j}(Q,T)}{h}
\left(e^{\frac{Q}{kT}}-(-1)^{2s_j}\right)^{-1}
\times\frac{{\rm d}g_{j\bar{D}}(Q,E,p_0)}{{\rm d}E}{\rm
d}Q
\end{equation}

where ${\rm d}g_{j\bar{D}}(Q,E,p_0)/{\rm d}E$ is the fragmentation 
function into
antideuterons evaluated with this coalescence model for a given
momentum $p_0$.\\

In any case this spectrum is, then, convoluted with the PBH mass spectrum \cite{Carr4}
assumed to be scaling as $M^2$ below $M_*\approx 5\times 10^{14}$g and as 
$M^{-2.5}$ above $M_*$  and normalised to the local density.

\section{Propagation scheme}

\begin{figure}
\begin{center}
\psfig{figure=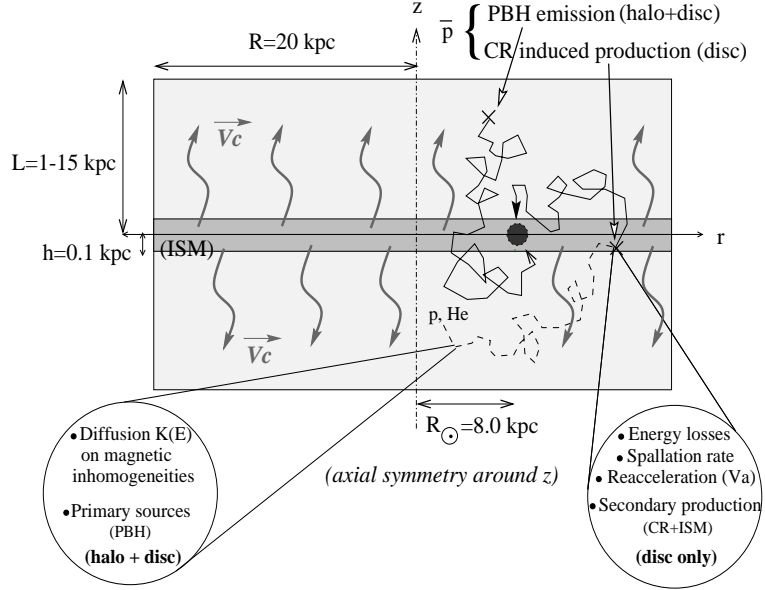,height=8cm,width=10cm}
\caption{
\label{diffusion}Schematic view of the axi-symmetric diffusion
model. Secondary antiproton sources originate from {\sc cr}/{\sc ism}
interaction in the disc only; primary sources
are also distributed in the dark halo which extends far beyond the diffusion
halo. Drawing by D. Maurin.}
\end{center}
\end{figure}

The propagation of cosmic rays throughout the Galaxy is described
with a refined two--zone
effective diffusion model which has been thoroughly discussed
elsewhere (Maurin {\it et al.} \cite{David}, Donato {\it et
al.} \cite{Fiorenza}).

The Milky--Way is pictured
as a thin gaseous disc with radius $R = 20$ kpc and thickness
$2 h = 200$ pc (see Fig. \ref{diffusion}) where charged nuclei
are accelerated and destroyed by collisions on the interstellar gas,
yielding secondary cosmic rays.
The thin ridge is sandwiched between two thick confinement layers of
height $L$, called {\em diffusion halo}.

The five parameters of this model are $K_0$, $\delta$,
describing the diffusion
coefficient $K(E) = K_0 \beta {\cal R}^{-\delta}$,  the halo
half-height $L$, the convective velocity $V_c$ and the Alfven
velocity $V_a$. Actually, a confident range for these five parameters has been obtained
by the analysis of cosmic ray data on charged stable nuclei \cite{David}. This
exhaustive study allows a fully consistant treatment of the problem.\\

The source distribution for PBHs was assumed to follow the usual isothermal halo
profile.

\section{Results}

\subsection{Upper limit on the PBH density with antiprotons}

\begin{figure}
\begin{center}
\psfig{figure=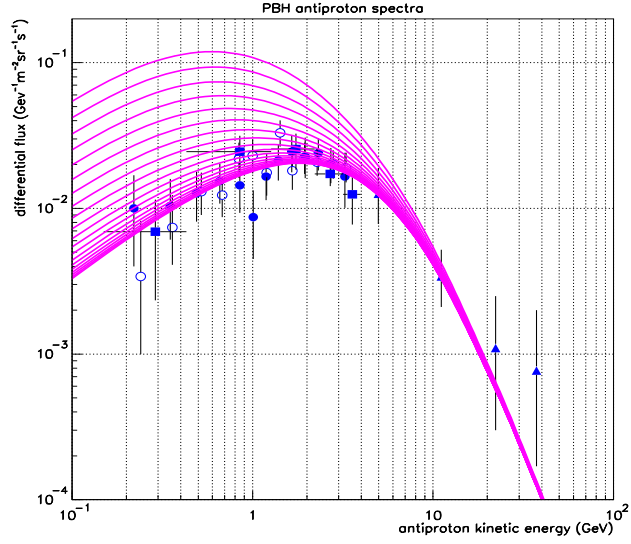,height=8cm,width=9cm}
\caption{Experimental data from BESS95 (filled circles), BESS98 (circles),
CAPRICE (triangles) and AMS (squares) superimposed
with mean theoretical {\sc pbh} spectra for $\rho_{\odot}^{PBH}$ between
$5\cdot10^{-35}~{\rm
g}.{\rm cm}^{-3}$ (lower curve) and $10^{-32}~{\rm g} \,{\rm cm}^{-3}$
(upper curves).
\label{pbar_tot}}
\end{center}
\end{figure}

Fig. \ref{pbar_tot} gives, for a fixed set of astrophysical parameters, the antiproton
flux due to the secondary and primary components \cite{Barrau4}. The lowest curve is without any PBH whereas
the upper one is for a local density $\rho_{\odot}^{PBH}=10^{-32}~{\rm g} \,{\rm cm}^{-3}$.
As expected, the experimental data can be reproduced without any new physics input. Taking
into account the statistical significance of the astrophysical uncertainties, an upper limit
on $\rho_{\odot}^{PBH}$ can be obtained \cite{Barrau4} as a function of the diffusion halo 
thickness $L$. For a "reasonable" value of this parameter around 3 kpc, the upper limit is 
$\rho^{PBH}_{\odot} < 5.3 \times 10^{-33}~{\rm g} \,{\rm cm}^{-3}$ which translates into
$\Omega_{PBH} \leq 10^{-8} \, \Omega_M \sim 4 \times 10^{-9}$ assuming that PBHs cluster as
dark matter.

\subsection{A new window for detection : antideuterons}

\begin{figure}
\begin{center}
\psfig{figure=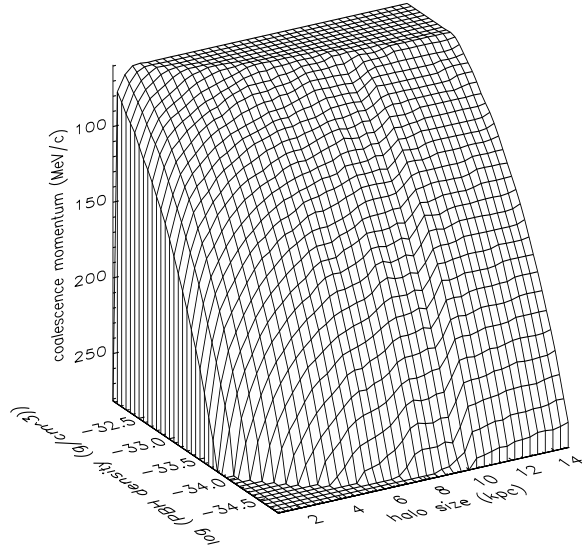,height=8cm,width=9cm}
\caption{Parameter space (halo thickness $L$ : 1-15 kpc ; coalescence momentum
$p_0$ : 60-285 MeV/c; {\sc pbh} density $\rho_{\odot}$ :
$10^{-35}-10^{-31}$g.cm$^{-3}$)
within the AMS sensitivity (3 years of data taking). The allowed region lies
below the surface.\label{fig:3d_ams}}
\end{center}
\end{figure}

To go beyond an upper limit and try to detect PBHs is seems very interesting to look for
antideuterons. Below a few GeV, there is nearly no background for kinematical reasons
\cite{fiorenza} and the possible signal due to PBHs evaporation could be easy to detect. We
have evaluated the possible range of detection for the AMS experiment \cite{Barrau5}. It is shown
on Fig. \ref{fig:3d_ams} as a function of the three unknown parameters  \cite{Barrau6}: $L$, $p_0$ and 
$\rho^{PBH}_{\odot}$. The sensitivity of the experiment should allow, for averaged
parameters, an improvement in the current best upper limit by a factor of six, if not a
positive detection.\\

A complete study of the uncertainties due to the PBHs halo profile, to the possible
photosphere near the event horizon, to the finite reheating temperature, to nuclear process and
experimental measurements can be found in Barrau {\it et al.} \cite{Barrau4} \cite{Barrau5}.
 
\section{Conclusion}

Primordial black holes have
been used to derive interesting limits on the scalar fluctuations
spectrum on very small scales
studies \cite{Kim2} \cite{Po}.
It was also found that {\sc pbh}s are a great probe of the
early Universe with a varying
gravitational constant \cite {Carr2}.
Significant progress has been made in the understanding of
the evaporation
mechanism itself, both at usual energies \cite{Parikh} and in the 
near-planckian
tail of the spectrum \cite{Barrau2} \cite{Stas2}. Looking for PBHs or improving the current
upper limit is therefore a great challenge for the forthcoming years.

\end{document}